\documentstyle[aps,prd,preprint]{revtex}
\newcommand{\be}{\begin{equation}}
\newcommand{\ee}{\end{equation}}
\newcommand{\ba}{\begin{eqnarray}}
\newcommand{\ea}{\end{eqnarray}}

\def\edth{\raise1.3ex\hbox{$\scriptscriptstyle/$}\mkern-9mu\partial}
\begin{document}

\title{Non-singular solutions of flux branes in M-theory and attractor
solutions}

\author{R. Bhattacharyya\footnote{E-address: rajsekhar@vsnl.net}} 
\address{ Department of Physics, Dinabandhu Andrews College, Calcutta 700084,
India}

\maketitle
\vspace{2.5cm}

\begin{abstract}

Here we present a class of solutions of M-theory flux branes, which are
non-singular at origin. These class of solutions help us to determine the field
strength at origin together with the behavior of it near origin. Further we
show a way to find the attractor solutions of such flux branes.


{PACS number(s): 11.25.-w  11.15.-q}
\end{abstract}

\newpage 
\section{Introduction}
     A magnetic flux tube is charecterised by the integral of field strength
over the transverse plane and the generalization of magnetic flux tubes to
higher rank and dimensions are known as flux branes or F$p$ branes. To be
precise an F$p$ brane in $D$-dimensions has $p+1$ dimensional Poincare
invarience in world volume together with $SO(D-p-1)$ rotational invariance
in $D-p-1$ transverse directions and a non zero rank $D-p-1$ field
strength tangent to the transverse dimensions.
      
      The example of a flux brane with gravity is the Melvin
universe[1], which is a F1 brane in 3+1 Einstein-Maxwell gravity and
can be embedded in string theory(there are also 
dilatonic generalization of F1 brane in 3+1 dimensions[2-4] in string
theory). The embedding of the
Melvin universe in M-theory and in string theory has created a lot of
inerest in particular 
in the type IIA F7 brane[5-7], whose M-theory description suggests IIA-0A 
duality [8,9] with a speculation that the end point of 0A
(closed string) 
tachyon condensetion is the IIA vacuum and F7-IIA cone
duality with an alternate IIA description [7,25-30].

       Flux branes also play an important role in supergravity description
of expansion of D$p$ brane into D$(p+2)$ and in the stabilisation of tubular
branes[10-12].Further the study of magnetic background allowing a conformal
theory description[13-16] gives a direction for studying the decay of unstable
background to stable supersymmetric ones[17-22] and when several
magnetic parameters are considered the supergravity background may preserve
some supersymmetry[23] and in the context of supersymmetric flux branes
their construction and classification has been analised extensively 
from supergravity persepective together with the search of dynamical
objects in the theory[24,42-46].

       There is an interesting conjecture[7] related to flux branes
and holographic principle[31,32], that the F$p$ branes in string theory
and M-theory are large N duals of field theories without gravity. This
conjecture is in the spirit of Ads/CFT[33] which is a realisation of
holographic principle.(There also exists analogue of holographic principle
at much larger length scale[34-36]). Attempts are made to understand such
non-supersymmetric holographic dualities studying solutions corresponding to
non-supersymmetric brane configurations, but the solutions turn out to be
 singular at origin[37-41].

      Here we find a non-singular solutions of F$p$ branes in M-theory. In
M-theory there are two types of F$p$ branes one is F6, a magnetic flux brane
and the other is F3, an electric one. Their solutions can be obtained from the
action in Einstein frame with graviton and $D-p-1$ form field for $D=11$ and for
$p=6$, $p=3$(here both dilaton and dilaton coupling are set at zero). In 
section 2 we construct a class of non singular solutions near origin but for
the sake of finite field strength we ultimately restrict ourselves to take
solutions upto origin. Further these new solutions near origin help to
determine the field strength at origin (in terms of the
parameters of the solutions) together with the behavior of the field near
origin.   

      In section 3 we extend the new solutions of section 2 with the
help of the solutions in[7] which are singular at origin i.e we will prove that
there exists a point at which the new solutions near origin and the
corresponding old singular solutions intersect. Thus we get solutions of
flux branes except at origin. 

      In sectin 4 we show that for non-singular solutions[7] which hold only 
near origin can be able to catch the corresponding solutions discussed in 
section 2.
Thus we present non-singular solutions of M-theory flux branes with large
field strength at origin. It is to be noted that the solutions in[7] i.e the 
non-singular
solutions near origin and the singular solutions, do not
intersect with each other.   

      In section 5 we propose a way to find the attractor solutions of
M-theory flux branes without solving the metric explicitly but assuming  
some tentative form of the solutions of metric. Further this method is
also applicable to determine the asymptotic solutions of II$q$ branes.(A
II$q$ flux brane is obtained from an action
in Einstein frame for $D=10$ where dilaton is present with graviton and $q$ form
field)
\section{Solutions near origin but not at origin}
As we have mentioned earlier that a F$p$-brane will have $p+1$ dimensional poincare
invarience in the world-volume and $SO(q)$ rotational invarience in 
$q=D-p-1$ transverse directions. For notational convience we will 
use the number of transverse dimensions $q$ to label the 
field strength of F$p$-branes.    

        So taking the graviton, a $q$ form field strength $F_q$ and 
in the case  of string theory a dilaton, the action in the Einstein frame is 
given by
$$S=\frac{1}{{l_p}^{D-2}}\int{d^D
x\surd{g}(R-\frac{1}{2}\partial\phi\partial\phi -\frac{1}{2q!} e^{a\phi}
{F_q}^2)} \eqno(2.1)$$    

        For M-theory we have $D=11$ and $\phi=a=0$, the field strength
 $F_q$
has either $q=4$ for a magnetic fluxbranes or the dual $q=7$ for an electric
fluxbranes.
Our ansatz for the metric is
$$ds^2 = e^{2A(r)} (-dt^2 +{dx_1}^2 +.....+d{x_{D-q-1}}^2) +dr^2 +e^{2C(r)}
d{{\Omega}_{q-1}}^2 \eqno(2.2)$$  
The above metric has the asymptotic form 
$$ds^2 \sim r^l (-dt^2 +d{x_1}^2 +.....+d{x_{D-q-1}}^2) +dr^2 +nr^2 
d{\Omega_{q-1}}^2 \eqno(2.3)$$	
 which is known as the attractor solution of the above metric, where $l$ and
$n$ are positive constants.
The equation of motion for the field strength $d*F=0$ can be solved as
$$F_q=fM_p e^{-(D-q)A+(q-1)C} \epsilon_{r{\alpha_1}...\alpha_{q-1}}
\eqno(2.4)$$ where $f$ is dimensionless constant which measures the field 
strength at the origin. Again for large $r$ the field strength behaves as 
$$F_q \sim sr^{q-2}\epsilon_q \eqno(2.5)$$
where s is a positive constants and $\epsilon_q$ is the
transverse volume.

So starting from above action we will seek for
solutions of M-theory flux branes together with a zero energy constraint,
 coming from the $R_{rr}$ component of Einstein's equation. For two fluxbranes
 here we first
consider the F6-brane, and from above action we arrive at
the following equations of motion			
$$A''+7A'A' +3A'C' -\frac{(fM_p)^2}{6} e^{-14A}=0 \eqno(2.6a)$$
$$C''+7A'C' +3C'C' +\frac{(fM_p)^2}{3} e^{-14A} -2 e^{-2C}=0 \eqno(2.6b)$$
The zero energy constraint implies
$$7A'A'+7A'C'+C'C'-\frac{(fM_p)^2}{12} e^{-14A} -e^{-2C}=0 \eqno(2.6c)$$
At the centre of the fluxbrane the metric is flat. So near origin we assume
$C=\ln{(\frac{4m}{3k}(\frac{14}{27})^{1/2}A)}$. So for $r\rightarrow 0$,
 $A\rightarrow 0$ implies $e^{2A} \rightarrow 1$ and $e^{2C}\rightarrow 0$,
 which satisfy the boundary condition of F6 brane. Now defining
$$A_2 (r)=\frac{kr^{3/2}}{(r+m)^{3/2}}
-\frac{3k}{2}(\frac{r}{m})^{1/2}(1+\frac{r}{m})^{1/2}
+\frac{3k}{2}(\ln{(\frac{r}{m})^{1/2}+(1+\frac{r}{m})^{1/2}}) \eqno(2.7)$$
where $1>m>0$ and $k>1$ and $A_2 \rightarrow 0$ for $r\rightarrow 0$ and 
combining equations of (2.6) into a single equation
$$A''(1+3A)+A'A'(2+7A)=0 \eqno(2.8)$$
it can be easily seen that for $A=A_2$ equation (2.8) holds near $r=0$ and
also it makes metric to be flat at origin. Here $A_2$ actually represents a
class of solutions where for each $m$ and $k$ in their corresponding ranges we
get each solution of the class and further for every m, $A_2$ remains a
solution of (2.8) as long as $\frac{r}{m}<1$

        Further one can find the field strength $fM_p$ at $r=0$ by using 
$A=A_2$ and $C=C_2 =\ln{(\frac{4m}{3k}(\frac{14}{27})^{1/2}A_2)}$, but it turns
out to be infinite for the cost of non-singularness of the solution at
origin. So for the sake of finite $fM_p$ we have to restrict ourselves to go
upto origin with this solution i.e we can allow a very large value of $fM_p$
and without any loss of generality using (2.6(a)) we can
write$$fM_p=(\frac{27k}{2m^2 \epsilon})^{1/2} \eqno(2.9)$$ where
$\epsilon=\frac{\bar{r}}{m}\not=0$ for some $r=\bar{r}$ together with the
field strength near origin by
$$F=(\frac{27 k}{2 m^2 \epsilon})^{1/2}e^{-7A_2+3C_2} \eqno(2.10)$$
Clearly our choices of $m$ and $k$ make field strength very large, but we fail to take (2.7) as a solution
upto origin, we can go as close as possible but we have to stop somewhere.
In other words the field strength at origin stops the solution at some
$r\not=0$ and for larger $fM_p$ we can get closer to origin with this
non-singular solution.
We can extend this solution upto origin and we will discuss it in section 4.   

     Now we cosider the F3 brane and so from (2.1)
 the equations of motion are 
$$A''+4A'A'+6A'C'-\frac{(fM_p)^2}{3}e^{-8A}=0 \eqno(2.11a)$$
$$C''+4A'C'+6C'C'+\frac{(fM_p)^2}{6} e^{-8A}-5e^{-2C}=0 \eqno(2.11b)$$ 

and the zero energy constraint becomes 
$$24A'A'+60C'C'+96A'C'-\frac{(fM_p)^2}{24} e^{-8A}-60e^{-2C}=0 \eqno(2.11c)$$

Here once again we want to find non-singular solution of $A$ together with
$C$.
So assuming $C=\ln{(\frac{4m}{3k}(\frac{20}{27})^{1/2}A)}$ and combining
 equations of (2.11) we get 
$$A''(4+3A)+2A'A'(2+4A)=0 \eqno(2.12)$$   
Again for $A=A_2$ of (2.7)
we find such $A$ satisfy (2.12) near origin, where  $C=C_2=
\ln{(\frac{4m}{3k}(\frac{20}{27})^{1/2}A_2)}$.
 Further we can find the field
strength at origin to be infinite so we restrict our solution not to
reach origin as we have done in F6 case, and also argueing previously we can
find the field strength at origin using (2.11(a)) 
$$fM_p=(\frac{27k}{2m^2 \epsilon})^{1/2} \eqno(2.13)$$ where
$\epsilon=\frac{\bar{r}}{m}\not=0$ for some $r=\bar{r}$
together with the behavior of $F$ near origin
$$F= (\frac{27k}{2 m^2 \epsilon})^{1/2} e^{-4A_2+6C_2} \eqno(2.14)$$
 We also extend this
solution upto origin in section 4. 
 
   So for F3 case field strength at origin has the same expression of that of
F6 but this does not necessarily imply that at origin both have the same value of field
strengths because $m$ and $k$ of (2.7) can take any value out of their
corresponding ranges. We will discuss it later.

\section{Attempt for solutions except at origin }
      In previous section we have got solutions of $A$ and $C$ of (2.2) for F6 
and F3 branes near
origin. This solutions are not valid for any $r$ specially for large $r$. This
can be verified by substituting $A$ and $C$ (as obtained in previous section) in
their equations of motions or by comparing to the attrctor solution, as every
 solutions asymptotically have to match with it. 

      We have discussed earlier that there exist solutions of metric of F6
and F3 branes i.e the solutions of $A$ which is singular at origin. So
here we can make an attempt to paste the solutions of $A$ given here with the
corresponding solutions in[7] such that we can get a solution that can be 
extended to any larger $r$ for both F6 and F3 branes. First we will do it for F6
brane.

      For F6 brane the singular $A$ and $C$ of (2.2) (call them $A_3$ and
$C_3$) are given by
$$A_3 =\frac{1}{7}\ln{(rfM_p)}-\frac{1}{14}\ln(18/7) \eqno(3.1a)$$ 
$$C_3 =\ln{r}-\frac{1}{2}\ln(27/14) \eqno(3.1b)$$
We will show that $A_3$ and $C_3$ intersect with $A_2$ and $C_2$
respectively. So we define $$f(r)=A_2 (r)-A_3 (r) $$
Clearly $f$ is continuous in $(0,\infty)$. Now for $r\rightarrow 0$ as
$A_3 (r)
\rightarrow -\infty$ and $A_2 (r)\rightarrow 0$, $f(r)\rightarrow \infty$. So
there exists $r_1$ very near to origin such that $f(r_1)>0$. 
      
      Now if we can
show that for some non zero $r_2$, $f(r_2)<0$ then by Bolzano's theorem of
continuous function[47] there exists some $r_0$ in $(r_1,r_2)$ where
$f(r_0)=0$ which implies $A_3 (r_0)=A_2 (r_0)$ and to do so we have to
carefull such that $r_0$ must not be much away from origin, and also
$\frac{r_0}{m}  <1$ otherwise $A_2 (r_0)$
will not be a solution there of (2.8).  
       
        So precisely we will show $f(r) <0$, for
$1>\epsilon_2 > \frac{r}{m}>\epsilon_1 >0$. We choose $k=2n$
where $n$ is positive integer
such that $\frac{3k}{2}(\epsilon_2)^{1/2}<1$.
Under such condition considering full expression of $A_2 (r)$ where for all
$\frac{r}{m} <1$ we have
$k\frac{(\frac{r}{m})^{3/2}}{(1+\frac{r}{m})^{3/2}} 
< \frac{3k}{2}
(\frac{r}{m})^{1/2}
(1+\frac{r}{m})^{1/2}$ and further there exists a constant $N$ such that 
$\ln{(N(\frac{3k}{2})^{2}(\frac{r}{m})^{1/2})}>\ln{((\frac{r}{m})^{1/2}+
(1+\frac{r}{m})^{1/2})^{3k/2}}$
So for $$g(r)=E+k\frac{(\frac{r}{m})^{3/2}}{(1+\frac{r}{m})^{3/2}}
-\frac{3k}{2}(\frac{r}{m})^{1/2}(1+\frac{r}{m})^{1/2} +\frac{5}{14}\ln{r}$$
where 
$$E=\ln{(N(3k/2)^2 (1/m)^{1/2}(18/7)^{1/14}(1/fM_p)^{1/7})}$$
we can choose large $fM_p$ or small $\epsilon$ such that $E\leq 0$
then $g(r_2)<0$
 for some $\epsilon_1 <\frac{r_2}{m}<\epsilon_2$ i.e 
$$f(r_2)=k\frac{(\frac{r_2}{m})^{3/2}}{(1+\frac{r_2}{m})^{3/2}}
-\frac{3k}{2}(\frac{r_2}{m})^{1/2}(1+\frac{r_2}{m})^{1/2}$$ $$+\ln{(\frac{((\frac{r_2}{m})^{1/2}+
(1+\frac{r_2}{m})^{1/2})^{3k/2}
(18/7)^{1/14}}
{(rfM_p)^{1/7}})}< g(r_2) <0$$
 Further we can always choose
$r_1 < r_2$ by continuity of $f$ and as $f(r)\rightarrow\infty$ for
$r\rightarrow 0$. So for very small $r_0$ in $(r_1,r_2)$,
$A_3 (r_0)=A_2 (r_0)$.

       Now $C_3 (r)$ and $C_2 (r)$ both matches near origin and we
can choose $r_0$ to be their point of pasting. So we define 
$${\cal A} (r)=A_2 (r)~~~~for~~r\leq r_0$$ $$=A_3 (r)~~~~for~~r\geq r_0$$   
$${\cal C} (r)=C_2 (r)~~~~for~~r\leq r_0$$ $$=C_3 (r)~~~~for~~r\geq r_0
\eqno(3.2)$$
Clearly these ${\cal A} (r)$ and ${\cal C} (r)$ are the solutions of
F6 brane metric except origin.

       In the same way we can also study F3. The singular solutions of it[7](here
$A_3 (r)$ and $C_3 (r)$ are the solutions of $A(r)$ and $C(r)$ of (2.2) )are
$$A_3 =\frac{1}{4}\ln{(rfM_p)}-\frac{1}{8}\ln(9/2) \eqno(3.3a)$$ 
$$C_3 =\ln{r}-\frac{1}{2}\ln(27/20) \eqno(3.3b)$$
 Again we can define a continuous
function $$f(r)=A_2 (r)-A_3 (r) \eqno(3.8)$$ and argueing in the same we have 
a very small non zero $r_1$ such that $f(r_1)>0$. Also under the same
codition i.e for
$1>\epsilon_2>\frac{r}{m}>\epsilon_1>0$ choosing $k$ to be an even positive 
integer such that for some, $\epsilon_1 <\frac{r_2}{m}<\epsilon_2$ 
$$f(r_2)=k\frac{(\frac{r_2}{m})^{3/2}}{(1+\frac{r_2}{m})^{3/2}}
-\frac{3k}{2}(\frac{r_2}{m})^{1/2}(1+\frac{r_2}{m})^{1/2}$$ $$ +\ln{(\frac{((\frac{r_2}{m})^{1/2}+
(1+\frac{r_2}{m})^{1/2} )^{3k/2}
(9/2)^{1/8}}
{(rfM_p)^{1/4}})} <0$$
and arguing in the previous way as we have done for F6 case 
 that there exists $r_2 >r_1 >0$ near origin,
and $f(r_2)$ changes its sign. So by Bolzano's theorem for some $r_0$ in
$(r_1,r_2)$, $A_2 (r_0) =A_3 (r_0)$.

       Here we can say the same thing for $C_2(r)$ and $C_3 (r)$  so we can define 
$${\cal A} (r)= A_2 (r)~~~~for~~r \leq r_0 $$  $$=A_3 (r)~~~~for~~r \geq r_0$$
$${\cal C} (r)=C_2 (r)~~~~for~~r\leq r_0$$ $$=C_3 (r)~~~~for~~r \geq r_0
\eqno(3.10)$$
Thus this ${\cal A} (r)$ and ${\cal C} (r)$ are again the
 solution of $A(r)$ and $C(r)$ of (2.2) for F3 brane except at origin.

\section{The non-singular solution }
     In section 2 we have already mentioned that if we restrict $f$ from
infinity we can not call (2.7) as a solution at origin or more precisely if we
take $\epsilon=\frac{\bar{r}}{m}$ then one can not carry (2.7) as a solution far
from $r$ towards origin.

       Let for $\bar{r}$, $fM_p$ is such, it manages $E\leq 0$, further
$\bar{r}$ is small
 so $A_2 (\bar{r})=\frac{3k\bar{r}}{4m}$. Now very close to
origin $$A_1 (r)=\frac{1}{48}(rfM_p)^2 \eqno(4.1)$$ is a solution[7] of
$A$ of 
(2.6) together with $C=C_1 =\ln{((\frac{14}{27})^{1/2}r)}$. So
at $r=\bar{r}$, $A_1 (\bar{r}) <A_2 (\bar{r})$ and for some 
$r>\bar{r}$, $A_1 (r) >A_2 (r)$, if not then $A_1 (r)$ must have an
intersection with $A_3 (r)$ of F6 but it is not so. So there exists $\bar{r_0}
< r_0$ such that $A_1 (\bar{r_0})=A_2 (\bar{r_0})$. 

        So using results of section 3 and with $A_1$, $A_2$, $A_3$ of F6 we
have
$$A(r)=A_1 (r)~~~~for~~r \leq \bar{r_0}$$ $$=A_2 (r)~~~~for~~r \leq r_0 $$  
$$=A_3 (r)~~~~for~~r \geq r_0 \eqno(4.2)$$
Similar things can be done with $C$ and with $C_1$, $C_2$, $C_3$ of F6. This
above A is the non-singular solution of F6.

       In the same way for F3 there also exists a solution of $A$
$$A_1 (r)=\frac{1}{42}(rfM_p)^2 \eqno(4.3)$$
near and upto origin with $C=C_1 =\ln{((\frac{20}{27})^{1/2}r)}$ 
 and in the same way we can show that there exists
$\bar{r_0}
< r_0$ near origin where $A_1$ and $A_2$ intersects. So again with $A_1$,
$A_2$, $A_3$ of F3 we
have non-singular $A$ of F3 can be defined like F6
and same thing can be done for $C$ of F3 also.

\section{A way to find attractor solution of M-theory flux branes}       
       In this section we attempt for a new way to find attractor solution
of the metric of F6 and F3(except constant $n$ see (2.3)) without solving
$A$ and $C$ of (2.2) explicitly. Looking at the
the equations of motion of flux branes one can find $A$ and $C$ have to blow up
for $r\rightarrow\infty$. So one of the choice out of the set of elementary
functions may be both $A$ and $C$ are like  
$r^a$ for some $0<a<1$ but then $A''$ falls faster than $A'$ and $C'$.
But if the choice of $A$ and $C$ are like $\ln{r}$ then $A''$, $A'A'$,
$A'C'$ and $C''$
all fall like $\frac{1}{r^2}$ for large $r$. So for
 constants $\alpha$ and $\beta$ we choose $A=\alpha\ln{r}$ and 
$C=\beta\ln{r}$. Under such condition (2.6) implies at large $r$,
both $(fM_p)^2 e^{-14A}$ and $e^{-2C}$ also behave as $\frac{1}{r^2}$
so  $$\alpha=\frac{1}{7}~~~~and~~~~ \beta=1 \eqno(5.1)$$
and thus for large $r$ we have 
$$ds^2 \sim r^{\frac{2}{7}}(-dt^2 +d{x_1}^2+.....+d{x_{6}}^2 )+dr^2
+r^2 d{\Omega_{q-1}}^2 \eqno(5.2)$$
and this is the attractor solution of F6 flux brane(except constant $n$).

       For F3 we can go through the same way, i.e taking
  $A=\alpha\ln{r}$ and $C=\beta\ln{r}$
and using (2.11) we get$$\alpha=\frac{1}{4}~~~~and~~~~ \beta=1 \eqno(5.3)$$
and asymptotically metric becomes
$$ds^2 \sim r^{\frac{1}{2}}(-dt^2 +d{x_1}^2+.....+d{x_{3}}^2 )+dr^2
+r^2 d{\Omega_{q-1}}^2 \eqno(5.4)$$
which is the attractor solution of F3 flux brane(except $n$).

     [Although here we only study the flux branes in M-theory but in this
context it is very
natural to discuss about the asymptotic solutions of type II$q$ flux branes, as
the way we have described above the asymptotic solutions of M-theory flux
branes also is a way for that of type II$q$ case. 

       In type II case there is dilaton in the action (2.1).  
For $D=10$ and the dilaton coupling $a=1/2(5-q)$, taking same ansatz (2.2) for
the metric and using the integral of
motion $\phi=4\frac{5-q}{q-1} A$ we have the equations of motion only for
$A$ and $C$ as
$$A''+(10-q)A'A'+(q-1)A'C'-\frac{q-1}{16}(fM_s)^2
e^{-2\frac{15+q}{q-1}A}=0 \eqno(5.5a)$$ $$C''+(10-q)C'A'+(q-1)C'C'-\frac{9-q}{16}
(fM_s)^2 e^{-2\frac{15+q}{q-1}A}-(q-2)e^{-2C}=0 \eqno(5.5b)$$  
together with the zero energy constraint 
$$[(10-q)(9-q)-8\frac{(5-q)^2}{(q-1)^2}]A'A'+(q-1)(q-2)C'C'+2(q-1)(10-q)A'C'$$
$$-(q-1)(q-2)e^{-2C} -\frac{1}{2}(fM_s)^2 e^{-2\frac{15+q}{q-1}A}=0 
\eqno(5.5c)$$

      Again if one
assume $A=\alpha\ln{r}$ and $C=\beta\ln{r}$ then in similar
manner we can have $$\alpha=\frac{q-1}{15+q}~~~~and~~~~ \beta=1 \eqno(5.6)$$ and for
large $r$ we have 
$$ds^2 \sim r^{2\frac{q-1}{15+q}}(-dt^2 +d{x_1}^2+.....+d{x_{9-q}}^2 )+dr^2
+r^2 d{\Omega_{q-1}}^2 \eqno(5.7)$$. This is the asymptotic solution of
metric of type
IIq case except the multiplicative constants of $r^{2\frac{q-1}{15+q}}$ and
$r^2$.]      

\section{Conclusion}

       Although the solutions $A(r)$ of flux branes as given in section 4 is
non-singular at origin but we have to show whether at $r_0$ and $\bar{r_0}$ 
derivatives of $A$ exist or not. However it can help
to find the value of $fM_p$ (or atleast a range of values) for F6 and F3 cases. 
      
     So to get an estimate of $fM_p$ we have to fix $m$ and $k$ of the solutins
or atleast to restrict their ranges further. To do so one can choose them in
such a way that (if possible) the difference of one sided derivatives of $A(r)$
both at $r_0$ and
$\bar{r_0}$ get minimised and otherwise we have to optimise. In this way we
can compare the values of $fM_p$ for F3 and F6 cases.

{\bf  Acknowledgement}\\
I am greatful to D.Gangopadhyay and K.Ghosh for useful discussion.

\end{document}